\documentstyle[12pt,epsf]{article}

\begin{document}
\baselineskip 18pt
\begin{flushright} 
DPNU-97-47 \\ 
October 1997 \\
hep-th/9710029 
\end{flushright} 

\vspace{5mm}

\begin{center}
{\huge\bf Matrix String Theory from \\
\vspace{3mm}
Brane Configuration}
\end{center}

\vspace{5mm}

\begin{center}
{\large\bf Kei Ito$^{1,2}$} and 
\underline{{\large\bf Nobuhito Maru$^2$}} 
\end{center}

\begin{center}
$^1$Department of Electrical and Computer Engineering \\
Nagoya Institute of Technology \\
Showa-ku, Nagoya 466, JAPAN \\
Tel: +81-52-735-5152, Fax: +81-52-735-5158
\end{center}

\begin{center}
$^2$Department of Physics, Nagoya University \\ 
Chikusa-ku, Nagoya 464-01, JAPAN \\
Tel: +81-52-789-2450, Fax: +81-52-789-2932 
\end{center}

\begin{center}
{\normalsize {\tt ito@ks.kyy.nitech.ac.jp}} \\ 
{\normalsize {\tt maru@eken.phys.nagoya-u.ac.jp}} 
\end{center}
\vspace{5mm}
\setcounter{page}{0}
\thispagestyle{empty}
\begin{center}
{\bf Abstract}
\end{center}
Configurations of fivebranes, twobranes and fourbranes 
in type IIA string theory, which give (1+1) dimensional 
supersymmetric gauge theories in the low energy limit, 
are constructed. 
It is shown that these brane configurations 
are equivalent to a certain class of matrix string theories. 
This opens up an avenue for the investigation of matrix string 
dynamics via the geometry of brane configurations 
in type IIA string theory. 
\vspace{5mm}
\begin{flushleft}
\end{flushleft}

\begin{flushleft}
%\begin{array}{rl}
\hspace{2.1cm}Supersymmetric gauge theories 
%\end{array}
\end{flushleft}

\newpage
Recently, there has been much progress in 
the investigation of the non-perturbative dynamics 
of three and four-dimensional supersymmetric Yang-Mills 
theories in the context of a ten-dimensional superstring 
theory and eleven-dimensional ``M theory". 
Among many important discoveries, the following two sets 
are remarkable. 
\begin{enumerate}
 \item A non-perturbative duality 
called ``mirror symmetry" 
of three-dimensional(d=3) supersymmetric 
Yang-Mills theory was explained by means of 
a certain brane configurations of 
Neveu-Schwartz(NS) type and Dirichlet(D)-type 
embbeded in 10dim type IIB superstring theory 
by Hanany and Witten \cite{HW}. 
A similar brane configurations in type IIA 
string theory which is responsible for dualities 
of four-dimensional(d=4) supersymmetric Yang-Mills 
theories (``Seiberg dualities") was found by 
Elitzur et al \cite{Elitzer}. 
This sort of brane configurations proved to be very 
powerful also in the investigation of the structure of 
Coulomb and Higgs branch of d=4 \cite{Witten1} 
and d=3 \cite{Boer} supersymmetric 
Yang-Mills theories. 
  \item A proposal was made by Banks et al. \cite{BFSS} 
that non-perturbative type II string dynamics 
(or M theory dynamics) can be captured by means of 
an appropriate large N limit of U(N) supersymmetric 
quantum machanics. 
This theory takes a form of a matrix theory which 
describes M theory dynamics and hence it is called 
M(atrix) model. 
This model was generalized to a so-called 
``matrix string theory" \cite{Motl}, \cite{BS}, \cite{DVV} 
which is U(N) supersymmetric 
(1+1) dimensional field theory i.e. a string theory. 
A matrix string theory in the presence of the 
longitudinal fivebranes was studied by Witten \cite{Witten}, 
in the context of (1+1) dimensional super Yang-Mills 
theory and was shown to be related to a moduli space of 
instantons of Yang-Mills theory on ${\bf R}^4$
\end{enumerate}

In this paper, we intend to unify these two approaches. 
To do so, we construct a new set of models with 
brane configurations embedded in type IIA 
string theory. The crucial difference from the models 
constricted so far, is that our model has two-dimensional 
spacetime(d=2). 
(In contrast to d=3 or d=4 in the previous models)

Surprisingly, our model turns out to be the same as 
a matrix string theory in the presence of longitudinal 
fivebranes. Thus, we find relationship between 
Hanany-Witten brane configurations with d=2 space-time 
(generalization of first approach) and 
matrix string theories (second approach).

This opens up a possibility of investigating complicated 
problems of non-perturbative dynamics of 4 dimensional 
Yang-Mills theories, such as moduli space of instantons, 
which has arisen in the second approach \cite{Witten}, 
by means of much simpler problems in the first approach 
in the geometry of corresponding 
brane configurations in superstring theories.

To show how our model works in the investigation of 
non-perturbative aspects of 4 dimensional theory, 
we study one-instanton moduli space of U(k) gauge theory 
by means of correponding brane configuration.

Now let us construct our model. 
First, we consider the following brane configuration in 
type IIA string theory in ten dimensional 
Mikowski space ($x^0, x^1, \cdots, x^9$). 
The left(right)-moving supercharges $Q_L(Q_R)$ satisfies 
the following conditions since the theory is non-chiral. 
\begin{eqnarray}
Q_L &=& \Gamma^0 \cdots \Gamma^9 Q_L, \\
Q_R &=& -\Gamma^0 \cdots \Gamma^9 Q_R. 
\end{eqnarray}
The configuration involves three kinds of branes which are: \\
(1) NS fivebranes with world volume ($x^0x^1x^2x^3x^4x^5$). 
These branes preserve supercharges of the form 
$\epsilon_LQ_L + \epsilon_RQ_R$, with 
\begin{eqnarray}
\label{NS}
\epsilon_L &=& \Gamma^0 \cdots \Gamma^5 \epsilon_L, \nonumber \\
\epsilon_R &=& \Gamma^0 \cdots \Gamma^5 \epsilon_R. 
\end{eqnarray}
(2) D fourbranes with world volume ($x^0x^1x^7x^8x^9$). 
These branes preserve supercharges satisfying 
\begin{equation}
\label{D4}
\epsilon_L = \Gamma^0 \Gamma^1 \Gamma^7 
\Gamma^8 \Gamma^9 \epsilon_R.
\end{equation}
(3) D twobranes with world volume ($x^0x^1x^6$) which preserve 
supercharges satisfying 
\begin{equation}
\label{D2}
\epsilon_L = \Gamma^0 \Gamma^1 \Gamma^6 \epsilon_R.
\end{equation}

Comparing the conditions of (\ref{NS}) and (\ref{D4}), 
we find that configurations of NS5 branes and D4 branes 
preserves one quarter of the original supersymmetries. 
The condition for the supercharges on D2 branes does not 
break the supersymmetries further. 
Therefore, one quarter of the original supersymmetries are 
preserved in the presence of these three kinds of branes.

We consider the dynamics on the world volume of n twobranes 
suspended between NS5 branes, 
and k fourbranes are placed at values of $x^6$ that are 
between the position of two NS fivebranes (Fig 1).

The world volume of D2 brane is three dimensional but 
it has only a finite length in $x^6$ direction 
(the distance between NS5 branes), so macroscopically, 
the twobrane dynamics is (1+1) dimensional. 
The original supersymmetry is N=2 in ten dimensions which 
is equivalent to N=16 in (1+1) dimensions. 
Since the brane configuration breaks one quarter of 
the supersymmetries, the unbroken supersymmetry is 
N=4 in (1+1) dimensions. Hence, this brane configuration 
describes a 1+1 dimensional N=(4,4) supersymmetric U(n) 
gauge theory with k hypermultiplets.

This model is not a matrix string theory as it stands, 
since a matrix string theory has N=(8,8) supersymmetry 
\cite{Motl}, \cite{BS}, \cite{DVV}. 
This discrepancy, however, can be avoided easily, 
if we compactify the model in the $x^6$ direction. 
Compactifying the model and identifying the two NS5 branes, 
we obtain a brane configuration which looks like 
that in Fig 2. 
This kind of brane configuration is not so peculiar 
as it looks at first sight and it was considered in 
d=3 \cite{Boer} and d=4 \cite{Witten1} contexts, and known that 
supersymmetry is enhanced.

In the present case, n D2 branes intersecting 
the NS5 brane produce a hypermultiplet in an adjoint 
representation of U(n). These hypermultiplet in the 
adjoint and singlet representation of U(n), 
together with U(n) vector multiplet form an N=(8,8) 
U(n) vector multiplet. 
This multiplet together with k hypermultiplet 
$H_{\alpha}(\alpha=1,\cdots,k)$ in the fundamental 
representation of U(n), coincide with the field content 
of the matrix string theory with k longitudinal fivebranes. 
(longitudinal fivebranes in M theory are seen as 
D4 brenes in type IIA string theory.)

In fact, there are further evidences to believe 
that our brane configuration is equivalent to 
matrix string theory in the presence of k 
longitudinal fivebranes. The latter theory is studied 
previously by Witten \cite{Witten} 
in the context of N=(4,4) d=2 
supersymmetric Yang-Mills theory obtained from 
N=1, d=6 theory by dimensional reduction. 
According to Ref. \cite{Witten}, 
the six dimensional theory has an 
SU(2) R symmetry K which acts trivially on the gauge 
fields  while the bosonic fields in the hypermultiplets 
transform with K=1/2. 
Upon dimensional reduction to 2 dimensions, 
six dimensional gauge fields split into a two dimensional 
gauge fields plus scalars $\phi_i(i=1,2,3,4)$. 
The dimensional reduction produces an extra SO(4) 
R symmetry groups, under which scalars $\phi_i$ 
transform as vectors and the scalars in 
the hypermultiplets are invariant.

In fact, these two R symmetries, SU(2) and SO(4) 
can be seen explicitly in our brane configuration. 
In our brane configuration in type IIA string, 
the Coulomb branch of the low energy effective U(n) 
gauge theory in (1+1) dimensions, is depicted in Fig 3. 
D2 brane can move in $x^2,x^3,x^4,x^5$ directions, 
since its endpoint should move in 
the world volume of NS5 brane, which is 
($x^0,x^1,x^2,x^3,x^4,x^5$). 
D2 brane motion has SO(4) symmetry corresponding 
to the rotation of the ($x^2,x^3,x^4,x^5$) coordinates 
of D2 brane position. 
This symmetry appears as R symmetry SO(4) in the low 
energy effective gauge theory in (1+1) dimensions. 
The D2 brane position ($x^2,x^3,x^4,x^5$) transform as 
a vector representation of the rotation group SO(4), 
which corresponding to the fact that scalars $\phi_i$ 
transform as vectors of SO(4) R symmetry. 
The D4 branes which represent matter hypermultiplets, 
are invariant under this rotation, which corresponds to 
the fact that hypermultiplets are invariant under 
this R transformation.

Now, in our brane configuration in type IIA 
string, the Higgs branch of the low energy 
effective U(n) gauge theory in (1+1) dimensions, 
is depicted in Fig 4. 
In this case, D2 brane break up on the world volume 
of D4 brane, which is ($x^0,x^1,x^7,x^8,x^9$). 
Therefore, D2 branes can move in ($x^7,x^8,x^9$) directions, 
and this motion has SO(3) symmetry, the double covering of 
which is SU(2). 
This SU(2) can be identified as the SU(2) R symmetry K 
in the (1+1) dimensional effective theory.

Thus, we have seen that our configuration 
in type IIA string (n D2 branes, k D4 branes 
and a NS5 brane) is equivalent to matrix string theory 
with gauge group U(n) in the presence of k longitudinal 
fivebranes. 
The latter theory was studied by Witten \cite{Witten}, 
in the context of (1+1) dimensional N=(4,4) supersymmetric 
U(n) gauge theories and it was shown that 
the Higgs branch is the moduli space of n-instantons 
of U(k) gauge theory on ${\bf R}^4$. 
Combining this fact with our results, 
it turns out that non-perturbative aspects of four 
dimensional gauge theories, such as the structure of 
instanton moduli space, can be determined by 
an investigation of Higgs branch of brane configurations 
of type IIA string considered here.

To show how our model works, in the investigation 
of instanton moduli space of four dimensional 
gauge theories, we can consider the simplest example, 
i.e. n=1 and an arbitrary k. 
The moduli space of one-instanton solution of U(k) 
gauge theory on ${\bf R}^4$ can be studied by this example. 
This is equivalent to the Higgs branch of (1+1) 
dimensional N=(4,4) supersymmetric U(1) gauge with 
k hypermutiplets.

The corresponding brane configuration of type IIA 
string is found, by our corresponce between 
brane configuration and matrix string. 
It is depicted in Fig 5. $x^7,x^8$ and $x^9$ values 
of the points on D4 branes, where D2 brane terminates, 
are related to the expectation values of scalars 
in the matter hypermultiplets. 
The matter hypermultiplets of U(1) gauge theory is 
equivalent, from the point of view of N=2 supersymmetry, 
to a pair of multiplets $A$ and $B$, 
with U(1) charge $+1$ and $-1$. 
For the D4 brane \#1, the relation between 
vacuum expectation value of the scalars 
in the corresponding hypermultiplet $A, B$ and 
$x^7,x^8,x^9$ values of the point where D2 brane comes in, 
is as follows: 
\begin{eqnarray}
x^7 &=& A_1\bar{A}_1 - B_1\bar{B}_1, \\
x^8+ix^9 &=& A_1B_1, \\
x^8-ix^9 &=& \bar{A}_1\bar{B}_1. 
\end{eqnarray}
For i-th D4 brane, the relation is found to be: 
\begin{eqnarray}
x^7 &=& \sum^i_{j=1} ( A_j\bar{A}_j - B_j\bar{B}_j ),  \\
x^8+ix^9 &=& \sum^i_{j=1} A_jB_j, \\
x^8-ix^9 &=& \sum^i_{j=1} \bar{A}_j\bar{B}_j. 
\end{eqnarray}
Due to the compactification in the $x^6$ direction, 
right-most D2 brane in Fig 5, is identified with 
left-most D2 brane.

The following conditions are imposed due to this fact. 
\begin{eqnarray}
\sum^k_{j=1} ( A_j\bar{A}_j - B_j\bar{B}_j ) &=& 0, \\
\sum^k_{j=1} A_jB_j &=& 0, \\
\sum^k_{j=1} \bar{A}_j\bar{B}_j &=& 0. 
\end{eqnarray}
A moduli space is a space of solutions of gauge theory 
modulo gauge equivalence. 
Therefore, $A_j, B_j$ differing by U(1) 
gauge transformation 
\begin{equation}
A_j \to e^{i\Lambda} A_j, B_j \to e^{-i\Lambda} B_j 
\end{equation}
should be identified. 
The D2 brane can detach from the NS5 brane 
and move in $x^7, x^8$ and $x^9$ directions. 
Then, the Neumann condition on the boundary of the 
D2 brane, which is $A_6=0$, $A_6$ being 
the sixth component of the gauge field, 
is relaxed and $A_6$ can have non-zero value. 
Thus, the moduli space of one-instanton solution of 
U(k) gauge theory on ${\bf R}^4$ is 
4k dimensional hyperk\"ahler manifold spanned 
by ($x^7, x^8, x^9$) values of D2-brane position 
$w^7, w^8, w^9$ and the value of $A_6$ together with 
$(A_j, B_j, \bar{A}_j, \bar{B}_j)$($j=1,\cdots,k$)
with the conditions (12)(13) and (14), 
with the identification (15). From 
the point of view of the low energy effective N=(4,4) 
supersymmetric gauge theory in (1+1) dimensions, 
the conditions (12)(13) and (14) are 
interpreted as vanishing 
of three components od D-terms in 
N=(4,4) supersymmetric 
gauge theory $D_0=0, D_+=0,D_-=0$. 
The conditions (12)(13) and (14) together with (15), 
coincides with the conditions derived 
by Witten \cite{Witten}, 
in the context of (1+1) dimensional N=(4,4) 
supersymmetric gauge theory.

This is just the first example to 
demonstrate how our model works in the 
investigation of instanton moduli space 
of gauge theories in four dimensions. 
There are many directions to pursue by means of 
our method. 
Inspecting the Higgs branch of the brane 
configuration which gives U(n) gauge theory 
with k hypermultiplets, we could determine 
the n-instanton moduli space of U(k) gauge 
theory on ${\bf R}^4$. 
The relationship between our brane configuration 
and ADHM construction of instanton moduli space 
would also be quite interesting to pursue. 
How the structure of Coulomb branch of (4,4) 
supersymmetric field theories in two dimensions 
studied in Ref. \cite{DS}, arises in our brane configuration 
might be another interesting problem. 
Works along these lines are now in progress.

Using our brane configurations, one can find a new fact on 
matrix string theory, which is so non trivial at the level of 
matrix string theory that the use of our brane configurations 
is indispensable. This is the duality (``mirror symmtery'') 
between matrix string theories.

Under T-duality transformation in $x^2$-direction to our branes, 
they are transformed as follows; D4 branes with world volume 
$(x^0,x^1,x^7,x^8,x^9)$ $\to$ D5 branes with world volume 
$(x^0,x^1,x^2,x^7,x^8,x^9)$, D2 branes with world volume 
$(x^0,x^1,x^6)$ $\to$ D3 branes with world volume 
$(x^0,x^1,x^2,x^6)$ while NS5 branes with world volume 
$(x^0,x^1,x^2,x^3,x^4,x^5)$ remain unchanged.

These branes are just the ingredients of the original 
Hanany-Witten brane configurations which give (2+1) dimensional 
gauge theories in the low energy limit.

In Hanany-Witten model \cite{HW}, there is an $SL(2,{\bf Z})$ symmetry 
or S-duality of the underlying type IIB string theory 
which results in the ``mirror symmetry'' of the effective 
three-dimensional gauge theories. 
Under this duality transformation, NS5 branes are transformed to 
D5 branes, and vice versa. Under this $SL(2,{\bf Z})$ 
transformation followed by R-transformation which exchanges 
$(x^3,x^4,x^5)$ with $(x^7,x^8,x^9)$. NS5 branes with wolrd volume 
$(x^0,x^1,x^2,x^3,x^4,x^5)$ are transformed to D5 branes with 
world volume $(x^0,x^1,x^2,x^7,x^8,x^9)$, and vice versa.

In order to go back to (1+1) dimensional matrix string theory, 
we perform T-duality transformation again. 
Then NS5 branes remain unchanged, whereas D5 branes are changed 
to D4 branes and D3 branes to D2 branes.

Therefore, under the successive application of T-duality, 
S-duality and T-duality transformation, the space-time dimension 
of effective gauge theory is transformed as follows
\begin{equation}
(1+1){\rm dim} \stackrel{T}{\longrightarrow} (2+1){\rm dim} 
\stackrel{S}{\longrightarrow} (2+1){\rm dim} 
\stackrel{T}{\longrightarrow} (1+1){\rm dim}
\end{equation}
and the branes are transformed as follows 
\begin{table}[h]
\begin{center}
    \begin{tabular}{ccccccc}
D4 & $\stackrel{T}{\longrightarrow}$ & D5 & 
$\stackrel{S}{\longrightarrow}$ & NS5 & $\stackrel{T}{\longrightarrow}$ 
& NS5 \\ 
NS5 & $\stackrel{T}{\longrightarrow}$ & NS5 & 
$\stackrel{S}{\longrightarrow}$ & D5 & $\stackrel{T}{\longrightarrow}$ 
& D4 \\ 
\end{tabular}
\end{center}
\end{table}

\noindent
In sum, D4 branes are transformed into NS5 branes, and NS5 branes 
are transformed into D4 branes under this TST-duality transformation.

Let us consider the brane configuration of $3$ NS5 branes, $2$ 
D4 branes and $n$ D2 branes of left-hand side of fig.6. 
The corresponding matrix string theory has a gauge group of $U(n)^3$. 
There matter hypermultiplets in the fundamental representations 
$(n,1,1)$ and $(1,n,1)$ of the gauge group and in the bifundamental 
representations $(n,\bar{n},1)$, $(1,n,\bar{n})$ and $(\bar{n},1,n)$.

Under the TST-duality transformation, D4 branes are changed to NS5 
branes, and vice versa and the resulting brane configuration 
in that of the right-hand side of fig.6. 
The corresponding matrix string theory is now a $U(n)^2$ gauge theory 
with matter hypermultiplets in the fundamental representations, 
$(n,1)$, $(1,n)$ and $(1,n)$ and bifundamental representations, 
$(n,\bar{n})$ and $(n,\bar{n})$.

Thus, we have found a pair of matrix string theories which are 
dual under TST-duality transformation. 
Finding this duality between this pair of matrix string theories 
without the use of brane configuration is quite difficult, 
if not impossible. Therefore, our correspondence relation 
between matrix string and brane configurations is indispensable 
in the study of matrix string theories.

Furthermore, we can find self-dual (self-mirror) model. 
Consider the configuration od fig.7, where the same number of 
NS5 and D5 branes are located in the order. The matrix string theory 
corresponding to the left-hand side is $U(n)^2$ gauge theory with 
hypermultiplets in the representations, $(n,1)$, $(n,1)$ and 
$(n,\bar{n})$, $(\bar{n},n)$. Exchanging D4 branes by NS5 branes, 
we obtain the brane configuration of the right-handed side, 
which is again the same matrix string theory. 
Therefore the model is self-dual.

Actually, essentially the same duality transformation as 
the TST-duality transformation considered here has been
applied to a simpler model by Brodie \cite{Brodie}. 
He applied it to $k$ 
D4 branes and no NS5 brane model, and obtained $k$ NS5 branes 
and no D4 brane. We have applied the same duality transformation 
to models where D4 branes and NS5 branes coexist. 
Therefore, we have generalized Brodie's dual pairs to more 
general and large class of pairs.

After completion of the original version of this work, 
we became aware of 
the papers \cite{Brodie},\cite{Iran} which 
overlap a part of our results. 
The discussion of the TST-duality transformation was added 
in revision.  
\begin{center}
{\bf Acknowledgements}
\end{center}
The authors would like to thank H. Ikemori and S. Kitakado for 
useful discussions. 
\newpage
\begin{center}
{\Large FIGURE CAPTIONS}
\end{center}
\begin{flushleft}
Fig 1: Brane configuration of D=2 N=(4,4) supersymmetric U(n)
Yang-Mills with k hypermultiplets. 
\end{flushleft}
\begin{flushleft}
Fig 2: Brane configuration of D=2 N=(4,4) 
supersymmetric U(n) Yang-Mills with adjoint, 
singlet, and k fundamental hypermultiplets. 
\end{flushleft}
\begin{flushleft}
Fig 3: Coulomb branch configuration of D=2 N=(4,4) 
supersymmetric U(n) Yang-Mills theory with adjoint 
and singlet hypermultiplets. 
\end{flushleft}
\begin{flushleft}
Fig 4: Higgs branch configuration of D=2 N=(4,4) 
supersymmetric U(n) Yang-Mills with adjoint, singlet, 
and k fundamental hypermultiplets
\end{flushleft}
\begin{flushleft}
Fig 5: Higgs branch configuration of D=2 N=(4,4) supersymmetric 
U(1) gauge theory with singlet and k matter hypermutiplets. 
\end{flushleft}
\begin{flushleft}
Fig 6: The left-hand side figure is brane configuration of 
supersymmetric $U(n)^3$ gauge thoery with $(n,1,1)$, $(1,n,1)$ 
fundamental hypermultiplets and $(n,\bar{n},1)$, $(1,n,\bar{n})$, 
$(\bar{n},1,n)$ bifundamental hypermultiplets. 
The right-hand side one is that of $U(n)^2$ gauge theory with 
$(n,1)$, $(1,n)$, $(1,n)$ fundamental hypermultiplets and 
$(n,\bar{n})$, $(\bar{n},n)$ bifundamental hypermultiplets. 
\end{flushleft}
\begin{flushleft}
Fig 7: Self-mirror brane configurations of supersymmetric $U(n)^2$ 
gauge theory with $(1,n)$, $(1,n)$ fundamental hypermultiplets and 
$(n,\bar{n})$, $(\bar{n},n)$ bifundamental hypermultiplets. 
\end{flushleft}
\newpage

\vspace*{1cm}
\begin{figure}
\begin{center}
\epsfbox{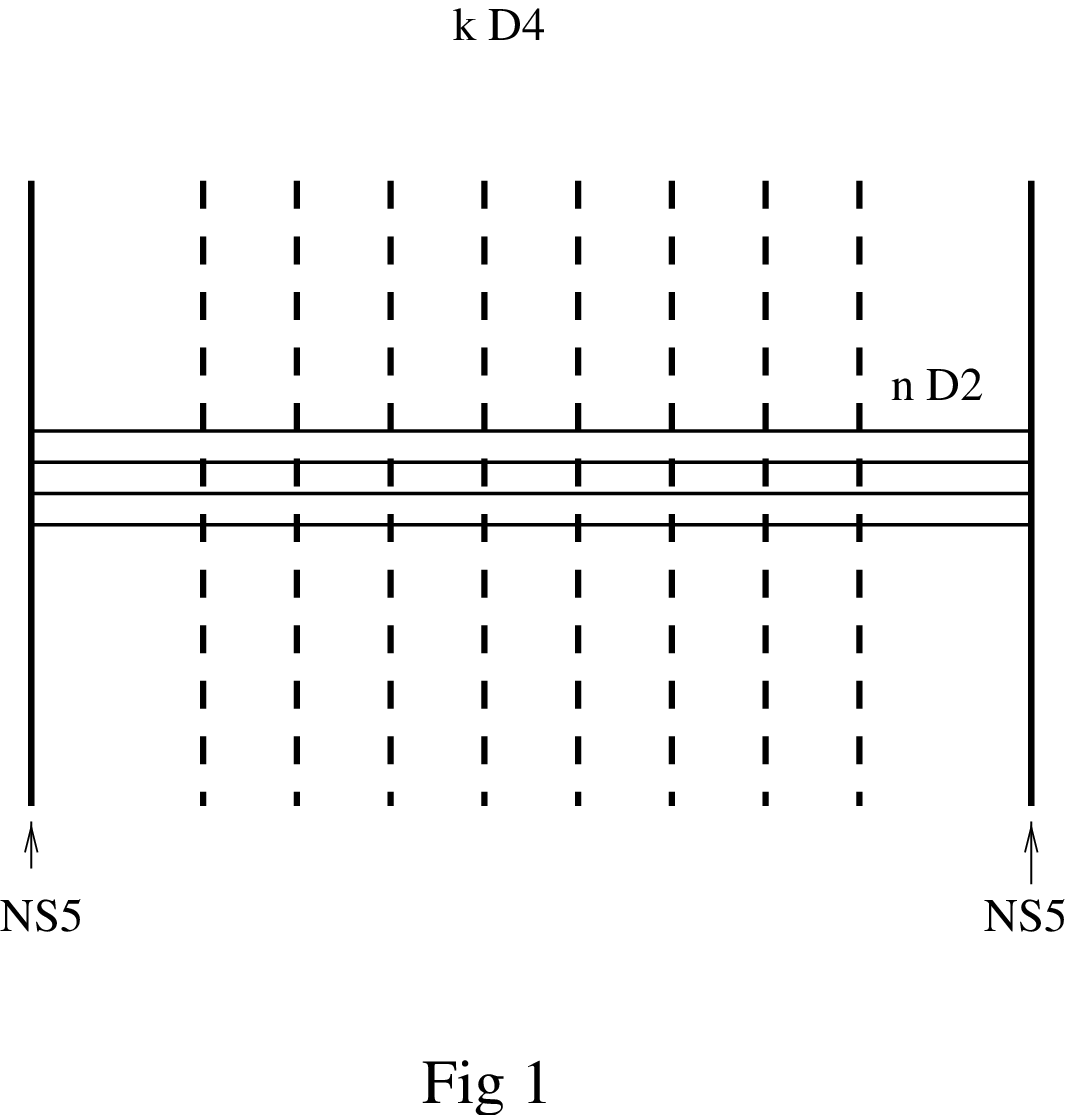}
\end{center}
\end{figure}

\newpage

\vspace*{1cm}
\begin{figure}
\begin{center}
\epsfbox{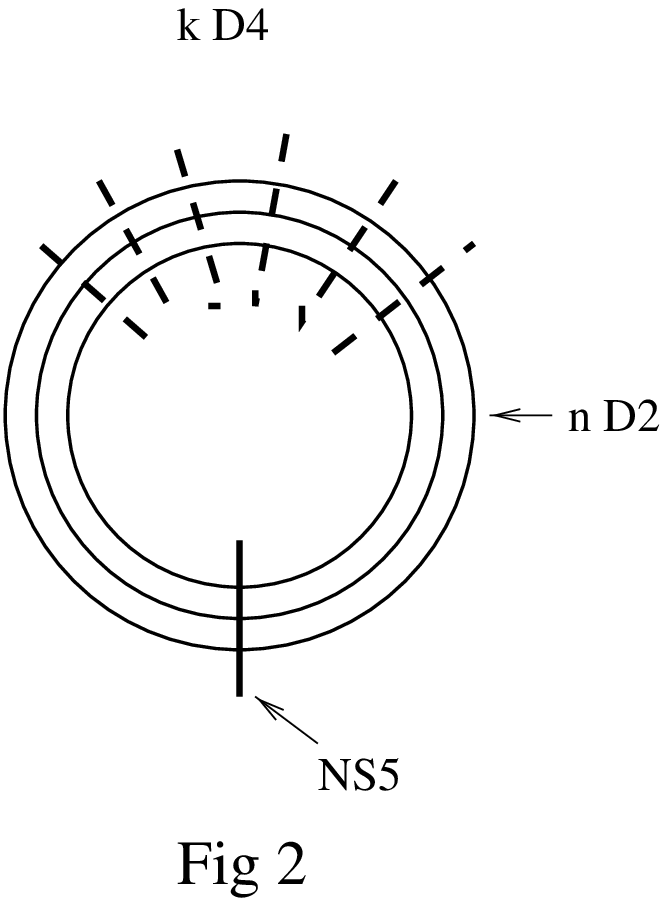}
\end{center}
\end{figure}

\newpage

\begin{figure}
\begin{center}
\epsfbox{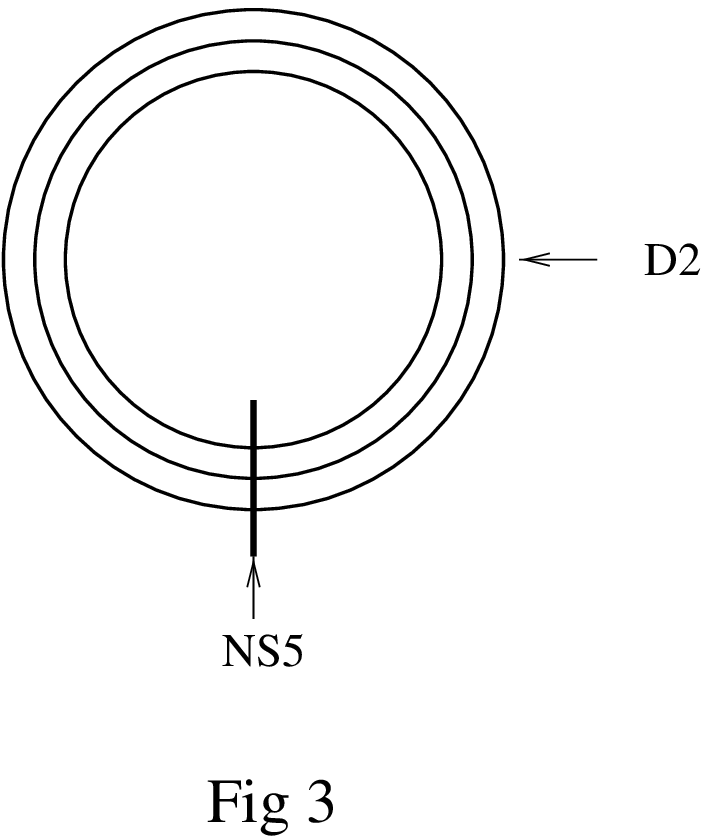}
\end{center}
\end{figure}

\newpage

\begin{figure}
\begin{center}
\epsfbox{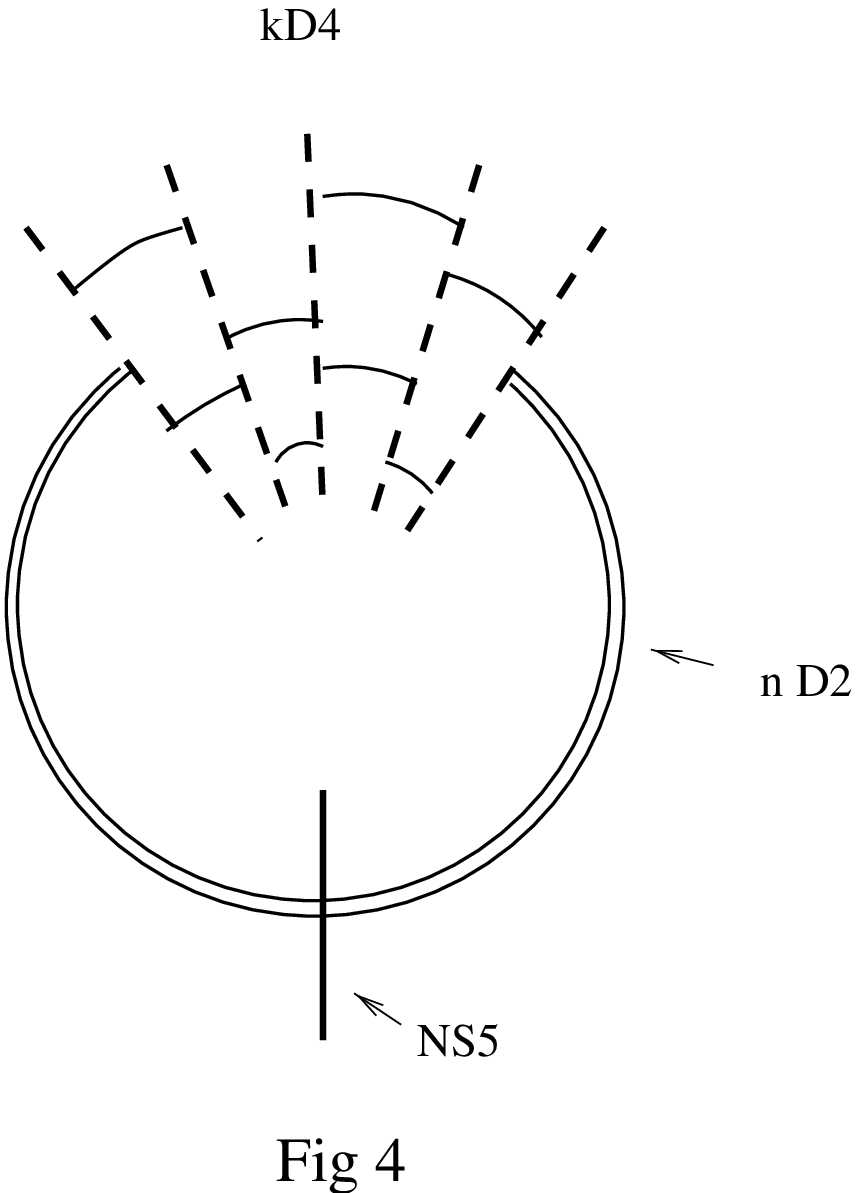}
\end{center}
\end{figure}

\newpage

\begin{figure}
\begin{center}
\epsfbox{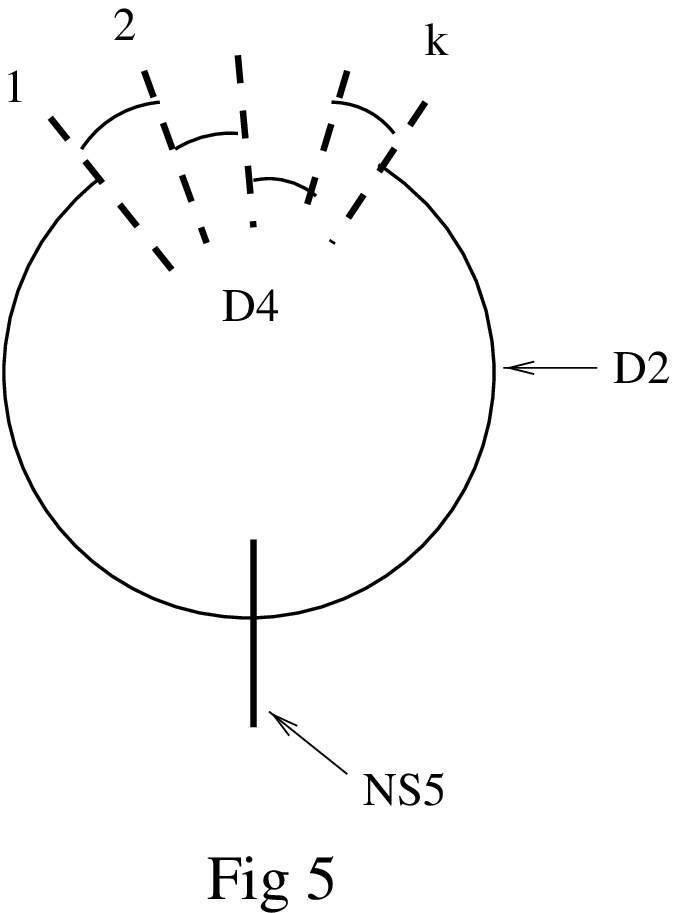}
\end{center}
\end{figure}

\newpage

\vspace*{3cm}
\begin{figure}
\begin{center}
\epsfxsize=16cm
\epsfbox{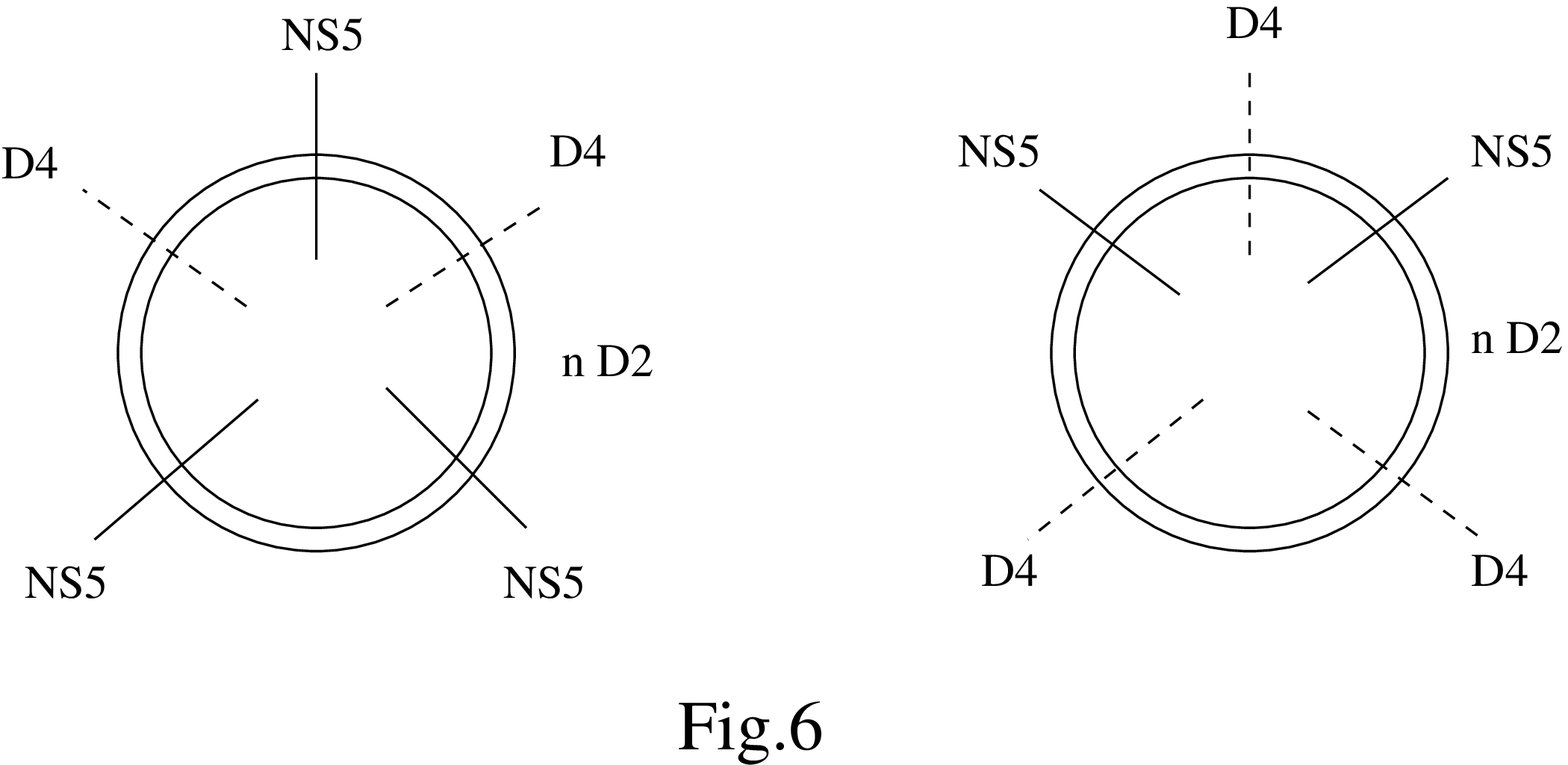}
\end{center}
\end{figure}

\newpage

\begin{figure}
\begin{center}
\epsfxsize=16cm
\epsfbox{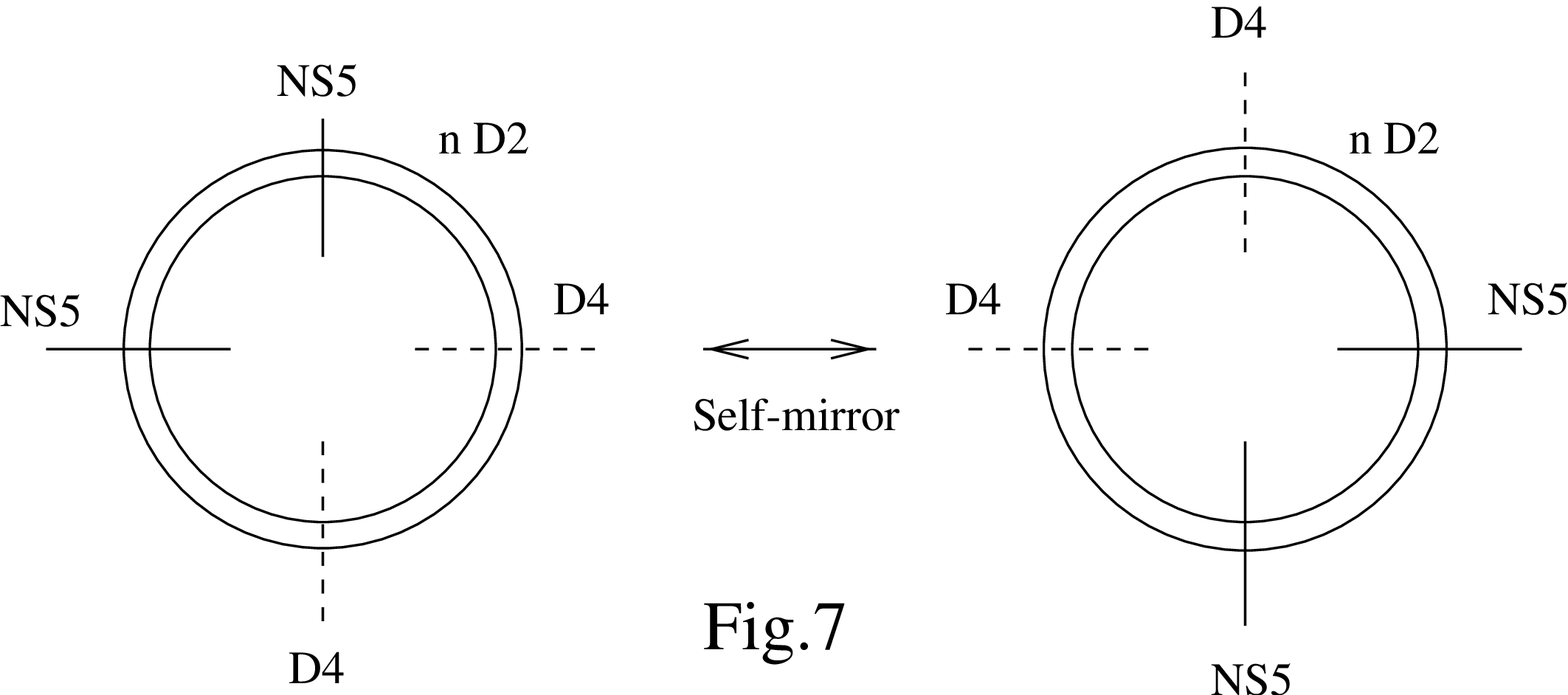}
\end{center}
\end{figure}

\end{document}